\documentclass{article}

\usepackage[final,nonatbib]{neurips_2019_ml4ad}

\usepackage[utf8]{inputenc} 
\usepackage[T1]{fontenc}    
\usepackage{hyperref}       
\usepackage{url}            
\usepackage{booktabs}       
\usepackage{amsfonts}       
\usepackage{nicefrac}       
\usepackage{microtype}      
\usepackage{graphicx}
\usepackage{subcaption}
\usepackage{xcolor}
\usepackage{amsmath}

\title{Unsupervised Neural Sensor Models for Synthetic LiDAR Data Augmentation}

\author{
  Ahmad El Sallab\thanks{The authors are ordered alphabetically} \\
  Valeo Egypt AI Research\\
  \texttt{ahmad.el-sallab@valeo.com} \\
  \And
  Ibrahim Sobh\protect\footnotemark[1] \\
  Valeo Egypt AI Research \\
  \texttt{ibrahim.sobh@valeo.com} \\
  \And
  Mohamed Zahran\protect\footnotemark[1] \\
  Valeo Egypt AI Research \\
  \texttt{mohamed.zahran@valeo.com} \\
  \And
  Mohamed Shawky\protect\footnotemark[1] \\
  Cairo University \\
  \texttt{mohamed.sabae99@eng-st.cu.edu.eg} \\
}

\begin{document}

\maketitle

\begin{abstract}
Data scarcity is a bottleneck to machine learning-based perception modules, usually tackled by augmenting real data with synthetic data from simulators. Realistic models of the vehicle perception sensors are hard to formulate in closed form, and at the same time, they require the existence of paired data to be learned. In this work, we propose two unsupervised neural sensor models based on unpaired domain translations with CycleGANs and Neural Style Transfer techniques. We employ CARLA as the simulation environment to obtain simulated LiDAR point clouds, together with their annotations for data augmentation, and we use KITTI dataset as the real LiDAR dataset from which we learn the realistic sensor model mapping. Moreover, we provide a framework for data augmentation and evaluation of the developed sensor models, through extrinsic object detection task evaluation using YOLO network adapted to provide oriented bounding boxes for LiDAR Bird-eye-View projected point clouds. Evaluation is performed on unseen real LiDAR frames from KITTI dataset, with different amounts of simulated data augmentation using the two proposed approaches, showing improvement of 6\% mAP for the object detection task, in favor of the augmenting LiDAR point clouds adapted with the proposed neural sensor models over the raw simulated LiDAR.

\end{abstract}

\section{Introduction}
\label{sec: introduction}
Machine learning has high potential to improve the perception module in the automated driving pipeline. Computer vision, enabled by machine learning and deep learning, provides high promises to improve environment perception from different sensors. However, machine learning models are known for their hunger to supervised data, which is expensive and time consuming to collect and annotate, creating an issue of data scarcity. The issue is amplified when dealing with non-camera sensors like LiDAR or Radar, where the physical nature of the signal is not common to human annotators, in addition to the high cost of the sensor itself in some cases and its limited availability and complex setup in other cases. 

Simulators are usually used to obtain data along with their free annotations provided by the simulator. The physics engines of simulators are developed to be photo-realistic, which requires to develop a camera model to produce the necessary visual effects for such realistic scenes, as if they are perceived by a camera in a real environment. The existence and maturity for similar models for other sensors like LiDAR is limited, or primitive to capture the real physical process that generates the real sensor data, and model the imperfect conditions in the real environment. 

In this paper, we tackle the problem of realistic sensor model for LiDAR. To model the imperfect real conditions, a mathematical model is needed to be developed for the generation process of the real LiDAR as perceived by the real sensor, which is hard to formulate in closed form and would have many assumptions.  In this work, a different strategy is adopted, which is to employ neural models to learn this mapping from simulation to 
real sensor data. Usually, a traditional paired data set is not available, with clean (simulated) frames and their corresponding noisy (real) frames, because this requires the existence of the sensor model in the first place. This makes traditional supervised optimization techniques not suitable for our task. Hence, we are proposing and evaluating two Neural Sensor Models (NSM) that do not suffer from this issue; CycleGANs and Neural Style Transfer (NST). For the first NSM, we formulate the problem of sensor modeling as an image-to-image translation and employ CycleGAN~\cite{Zhu-ICCV-2017} to learn the mapping from the simulated to real LiDAR. The second NSM is based on style transfer techniques using neural nets, where the content is the simulated LiDAR, while the style is the real LiDAR.

\begin{figure}
  \centering
  \begin{subfigure}[b]{.18\linewidth}
    \centering
    \includegraphics[width=\linewidth]{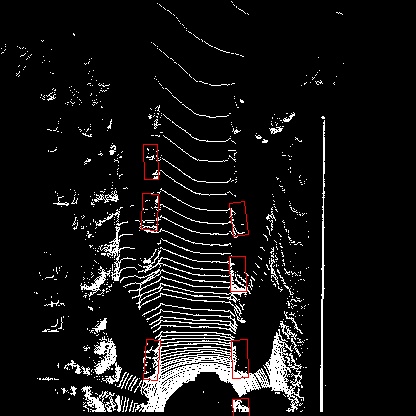}  
  \end{subfigure}
  \begin{subfigure}[b]{.18\linewidth}
    \centering
    \includegraphics[width=\linewidth]{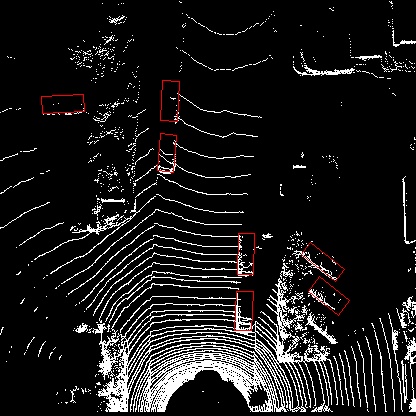}  
  \end{subfigure}
  \begin{subfigure}[b]{.18\linewidth}
    \centering
    \includegraphics[width=\linewidth]{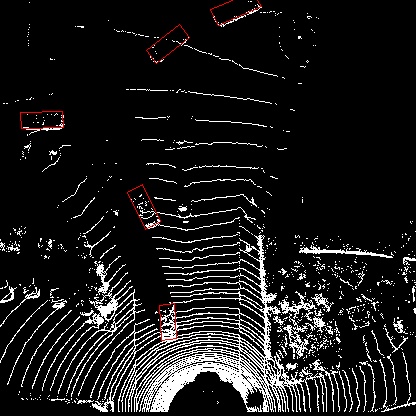} 
  \end{subfigure}
  \begin{subfigure}[b]{.18\linewidth}
    \centering
    \includegraphics[width=\linewidth]{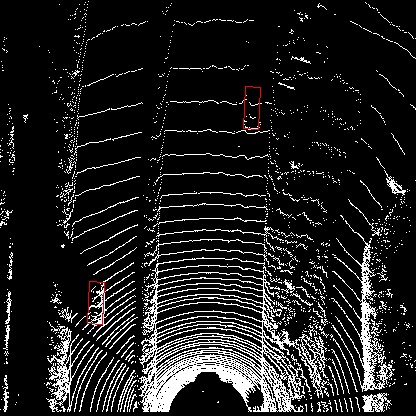} 
  \end{subfigure}\\
  \begin{subfigure}[b]{.18\linewidth}
    \centering
    \includegraphics[width=\linewidth]{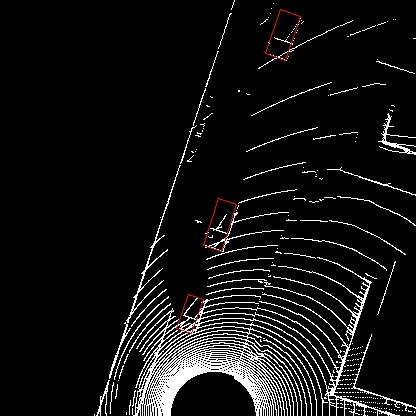}  
  \end{subfigure}
  \begin{subfigure}[b]{.18\linewidth}
    \centering
    \includegraphics[width=\linewidth]{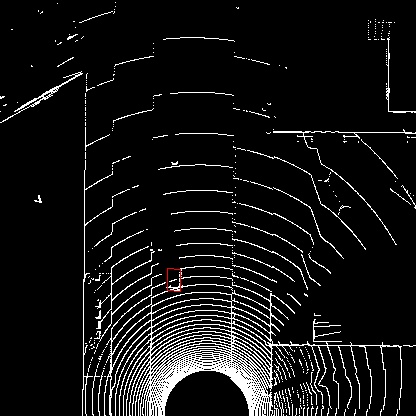}  
  \end{subfigure}
  \begin{subfigure}[b]{.18\linewidth}
    \centering
    \includegraphics[width=\linewidth]{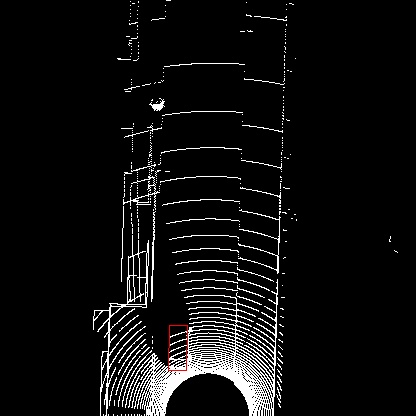} 
  \end{subfigure}
  \begin{subfigure}[b]{.18\linewidth}
    \centering
    \includegraphics[width=\linewidth]{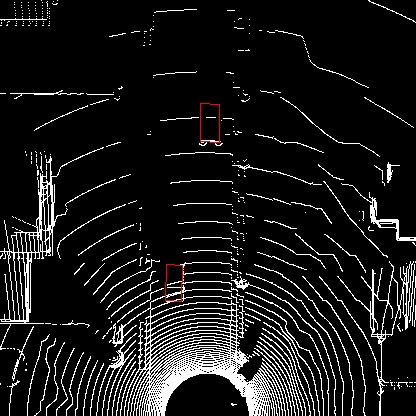} 
  \end{subfigure}
  \caption{BEV samples with ground truth bounding boxes. First row: KITTI dataset. Second row: CARLA simulator}
  \label{fig:samples}
\end{figure}

Sensors models are not perfect and could have artifacts, so we cannot depend only on simulated data to train machine learning models. Hence, simulated data is usually used for data augmentation purposes, where they augment real data to enrich the training data set. As the sensor model is more realistic, the percentage of simulated to real data in the augmented dataset can increase, which improves the training quality and reduces the collection and annotation cost. In this paper, we are proposing a framework for data augmentation from simulated LiDAR data, using the proposed sensor models. In our setup, we use CARLA \cite{DBLP:journals/corr/abs-1711-03938} as the simulation environment that provides the clean ray-tracing LiDAR PCL, together with its object bounding box annotations, while we use KITTI \cite{geiger2013vision}, as the real LiDAR data that we augment, and in the same time, use it to learn the sensor model mapping.  We choose object detection as the perception task in our augmentation framework, where we train a one-shot object detector, based on YOLO3D \cite{ali2018yolo3d} approach, modified to predict the orientation of the object bounding boxes of the projected LiDAR Point-Cloud (PCL) in the Bird-eye-View (BEV) space, which we call oriented YOLO. While the augmentation process involves simulated data, the evaluation test dataset only involves real LiDAR frames coming from KITTI dataset, with their human annotations of the bounding boxes. LiDAR BEV samples for CARLA and KITTI including ground truth bounding boxes are shown in figure \ref{fig:samples}.  

The data augmentation framework serves another purpose, which is the quantitative evaluation of the quality of the developed sensor model, rather than depending merely on the qualitative evaluation of the generated LiDAR with the expert human eye. We evaluate both models at different amounts of synthetic data augmentation, and compare their performance to the baseline of augmentation with raw simulated data without any sensor model mapping to show the advantage of employing a realistic sensor model.

The contributions of this work can be summarized as follows:
\begin{itemize}
  \item LiDAR sensor model based on CycleGANs.
  \item LiDAR sensor model based on NST.
  \item Evaluation of both sensor models at different amounts of data augmentation, through the proposed data augmentation framework to improve object detection from 2D BEV LiDAR.
\end{itemize}

The rest of the paper is organized as follows: first, the related work is presented, followed by the data augmentation framework and the details of the proposed sensor models. Following that, the experimental setup and the different environments and parameters are discussed for different augmentation scenarios and mappings from simulated CARLA to real KITTI data, evaluated by the mean Average Precision (mAP) metric of the object detection performance of oriented YOLO on real KITTI data.

\section{Related work}
\label{sec:related}
Domain adaptation is an important application of the generative models based on neural networks optimizations. Generative Adverserial Networks (GANs) are recently used in \textbf{image-to-image translation}, where the model is trained to map from images in a source domain to images in a target domain where conditional GANs (CGANs) \cite{DBLP:journals/corr/MirzaO14} are adopted to condition the generator and the discriminator on prior knowledge. Generally speaking, image-to-image translation can be either supervised or unsupervised. In the supervised setting, Pix2Pix~\cite{Isola-CVPR-2017}, SRGAN~\cite{ledig2017photo}, the training data is organized in pairs of input and the corresponding output samples. However, in some cases the paired training data is not available and only unpaired data is available. In this case the mapping is learned in an unsupervised way given unpaired data, as in CycleGAN~\cite{Zhu-ICCV-2017}, UNIT~\cite{liu2017unsupervised}. Moreover, instead of generating one image, it is possible to generate multiple images based on a single source image in multi-modal image translation. Multimodal translation can be paired Pix2PixHD~\cite{wang2018high}, BicycleGAN~\cite{zhu2017toward}, or unpaired MUNIT~\cite{huang2018multimodal}. CycleGAN is used in this work, as it performs conditional image-to-image translation on unpaired data.

\textbf{Neural Style Transfer (NST)} is another approach for domain adaptation.  The vanilla NST approach is based on Neural Algorithm of Artistic Style \cite{DBLP:journals/corr/GatysEB15a}, which is an online learning architecture, where the network optimizes both content and style loss for a single style image and a single content image. The vanilla NST algorithm is slow, as it runs an optimization process with every new generated image. To tackle the issue of slowness, several approaches exist in literature based on offline training. In \cite{DBLP:journals/corr/abs-1802-06474} an encoder is pre-trained on some auxiliary task, like a classification problem, and then both content and style images are passed to a pre-trained decoder using reconstruction and perceptual losses. Another approach in \cite{dumoulin2016learned} is comprised of an auto-encoder that takes the content image and multiple styles to produce the stylized image. We adopt the later approach, since it can be trained on multiple styles, also this architecture is significantly fast at inference.

\section{Neural sensor models for LiDAR data augmentation from simulated data}
\label{sec: ap}
\begin{figure}[ht]
    \vskip 0.2in
    \begin{center}
        \fbox{\includegraphics[scale=0.37]{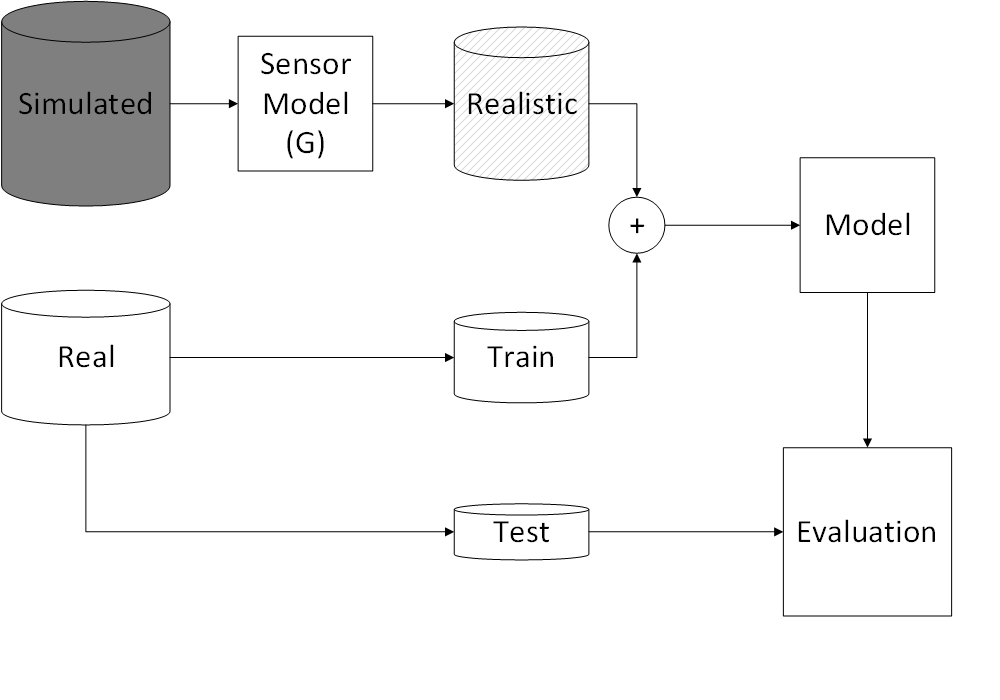}}
        \caption{Data augmentation framework}
        \label{fig:data_aug}
    \end{center}
\end{figure}

In this section, the proposed framework for LiDAR data augmentation from simulators is shown in figure \ref{fig:data_aug}, together with the two proposed sensor models. The simulated LiDAR PCL is obtained from the simulation environment, which is CARLA in our case. Then, the the simulated data is mapped through the sensor model mapping function, $G$, to obtain a realistic PCL, which is then added to the KITTI real LiDAR data. The augmented dataset now has LiDAR BEV frames, in addition to their corresponding ground truth annotations, either from human annotations in case of KITTI, or from the simulation environment in case of CARLA. The augmented dataset is then used to train an extrinsic evaluation model. We choose object detection as the evaluation task,  for which we use oriented YOLO adapted from YOLO3D \cite{ali2018yolo3d}, to produce oriented bounding boxes in the 2D BEV instead of predicting the 3D bounding box parameters. In fact, the only modification over YOLO3D is that we do not regress over the height or the z dimension in equation 9 in \cite{ali2018yolo3d}. The trained model is evaluated on unseen test set coming from real KITTI data, against the human annotated ground truth bounding boxes using the mAP metric.

The mapping function $G$, or the sensor model, can take different forms. In this work we evaluate three mappings:
\begin{itemize}
  \item $G=I$:  which means the raw simulated PCL is used for augmentation without any sensor model mapping. This acts as our baseline.
  \item $G=G_{cyc}$:  the CycleGAN sensor model.
  \item $G=G_{NST}$: the NST sensor model.
\end{itemize}

In the following sections, the details of the two proposed sensor models are described.

\subsection{CycleGAN}
\begin{figure}[ht]
    \vskip 0.2in
    \begin{center}
        \fbox{\includegraphics[scale=0.185]{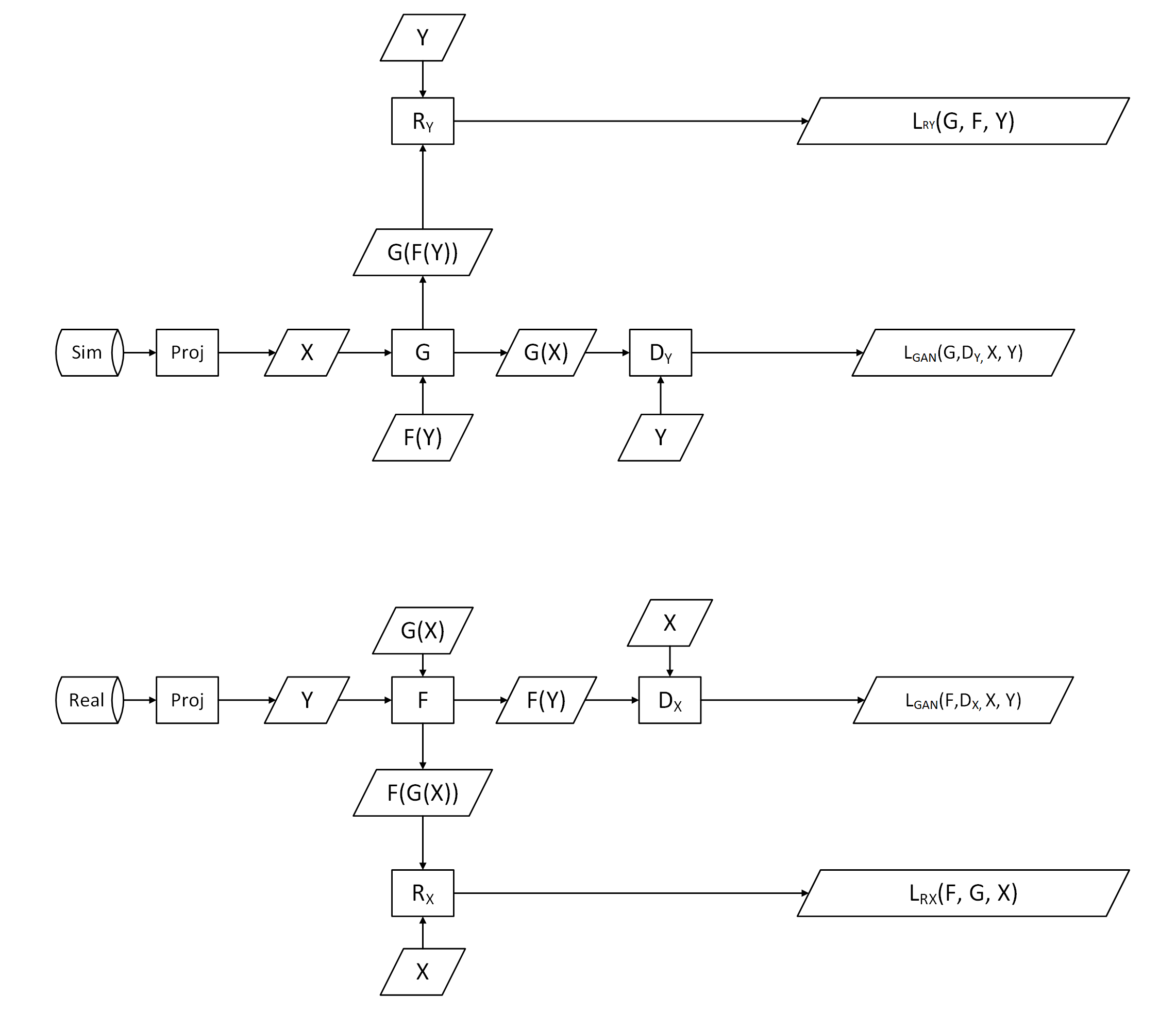}}
        \caption{CycleGAN model}
        \label{fig:vanilla_gan}
    \end{center}
\end{figure}

The CycleGAN NSM is shown in figure \ref{fig:vanilla_gan}. Let $X$ be the simulated data, and $Y$ be the real data. Both signals correspond to the raw $3$D PCL projected on $2$D space, through BEV projection. The forward generation network from $X$ to $Y$ (sim2real) is denoted $G$, which represents our desired sensor model. In the CycleGAN approach, another backward network (real2sim) form $Y$ to $X$ is defined and denoted as $F$.
Two adversarial losses are defined ${L}_{GAN}(G, D_Y, X,Y)$ and ${L}_{GAN}(F, D_X, X,Y)$ for both the forward and backward generators, with two different discriminators $D_Y$ and $D_X$. The objective of these generators is to fool the discriminators as in the normal GAN losses.

The main trick that enables CycleGANs to do the translation from unpaired data is the cycle consistency loss. Two cycle consistency losses are defined as follows ${L}_{R_Y}(G, F, Y)$ and ${L}_{R_X}(F, G, X)$, which aim at minimizing the $L1$ loss between the reconstructed forward or backward mappings; $G$ and $F$, and the original simulated or real data; $X$ or $Y$. Also, this is the main reason behind choosing CycleGAN over other GAN models.

The overall loss function is as follows:
\begin{eqnarray}
\mathcal{L}(G,F,D_X,D_Y)  =
    \mathcal{L}_{GAN}(G, D_Y, X,Y) \nonumber\\ 
 + \ \mathcal{L}_{GAN}(F, D_X, X,Y) \nonumber\\ 
    + \ \lambda \Big[\mathcal{L}_{R_Y}(G, F, Y) + \mathcal{L}_{R_X}(F, G, X)\Big] \;, \nonumber \\ 
\end{eqnarray}

The $\lambda$ parameter weights the cycle consistency versus the adversarial losses. The $G$ and $F$ mapping functions are obtained by optimizing the overall loss as follows:

\begin{eqnarray}
G^*, F^* = \arg \min_{G, F} \max_{D_X, D_Y} \mathcal{L}(G,F,D_X,D_Y)
\end{eqnarray}

\subsection{Neural style transfer}
\textbf{Style choice for $G_{NST}$}: We adopt the multiple style transfer architecture in \cite{dumoulin2016learned}, since we do not have a single representative style for the real data domain. In the CycleGAN image-to-image translation, $G_{cyc}$, we had a direct sim2real mapping, where we have the synthetic frame as input, and obtain the corresponding realistic frame. On the other hand, the NST generator $G_{NST}$ requires two inputs; the content, which is the synthetic frame, and the style which should come from the real data distribution.  During training, we choose $N$ styles by visually inspecting the real data (more details on that in the experimental section). During inference, we can only have one style as input, so each of the $N$ styles will produce a different sensor model $G_{NST}$. To choose one of the $N$ styles, we employ an extrinsic evaluation task to select the best performing style. For that, we train oriented YOLO only on real KITTI data to evaluate the mapped data using each $G_{NST}$ of the $N$ styles, and then we pick the best performing one. 2D bird-view is used as a LiDAR representation, where the 3D point cloud is projected into 2D representation without the height information. 

\textbf{NST generation time}: Another issue with vanilla NST \cite{DBLP:journals/corr/GatysEB15a} is that it is impractically slow, taking 5$min$ to generate a single frame on Nvidia Titan X$p$ GPU, due to the online optimization run with every frame. The classical trick of offline training of the network model is part of the adopted approach in \cite{dumoulin2016learned}, which reduces the generation time per frame to only 72$ms$ on the same GPU. As shown in figure \ref{fig:NST}, the modifications applied to the original architecture \cite{dumoulin2016learned} can be summarized as follows:   

\begin{figure}
  \centering
  \fbox{\includegraphics[width=\linewidth]{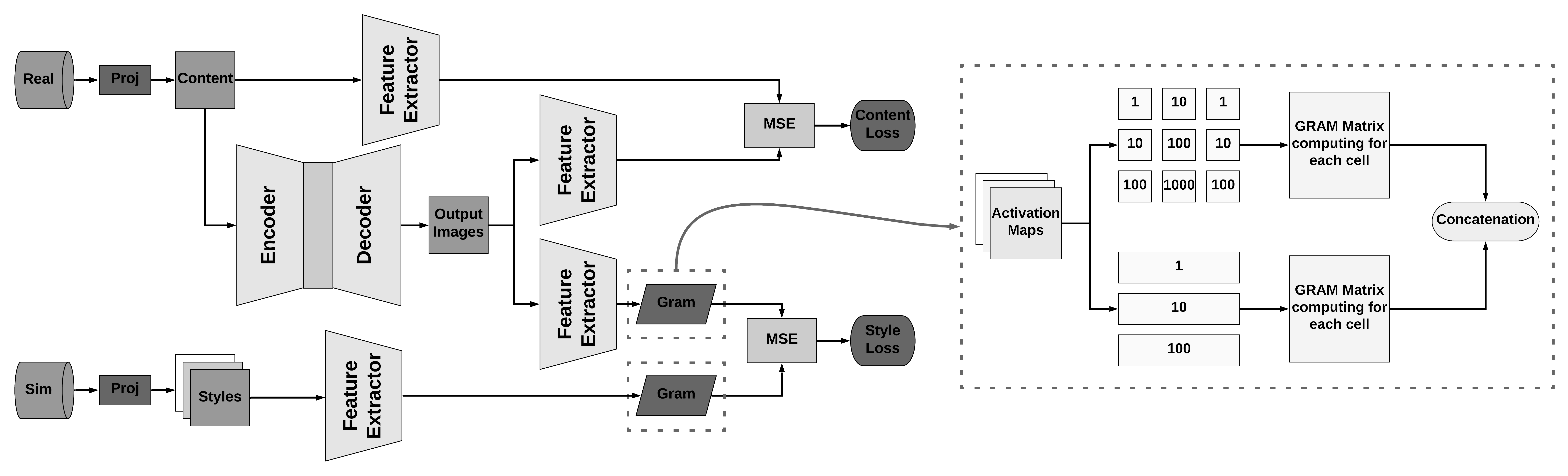}}
  \caption{Neural style transfer model}
  \label{fig:NST}
\end{figure}
  
\textbf{Feature extractor}: We noticed that the original feature extractor (VGG) is generic and huge, while LiDAR BEV data is relatively much simpler and more sparse, which does not require such architecture. Accordingly, VGG is replaced with a smaller encoder network of 4 convolution layers, each followed by a batch normalization layer, followed by a max pooling layer.

\textbf{Gram loss improvements}: The problem with the LiDAR BEV data is that they are sparse, thus it is hard to capture their style using Gram matrices. To overcome this issue, we apply localized style instead of applying the same style globally all over the content. Instead of comparing the Gram matrices of the whole activation maps, the activation maps are divided into $3 \times 3$ grid and the Gram matrix is computed for each cell independently. Then, the Gram matrices of the cells are compared to the corresponding ones of the style activation maps. For further improvement, an additional loss is added based on dividing the activation maps into horizontal slices and computing the Gram matrix for each slice. Moreover, different cells and slices are weighted empirically to increase the penalty of dense regions with higher importance. The Gram calculation part in figure \ref{fig:NST} shows the weights used for both cells and slices.

\textbf{Overall loss formulation}: 
 The loss of the NST network is formulated as follows:

\begin{equation}
\mathcal{L}(s,c,p)=\lambda_s\mathcal{L}_s(p)+\lambda_c\mathcal{L}_c(p)
\end{equation}

where $p$ represents the parameters of the transformer network, which takes the content and generate the stylized image, $ \mathcal{L}_s(p) $ is the style loss, $ \mathcal{L}_c(p) $ is the content loss and $ \lambda_s $, $ \lambda_c $ are scaling hyperparameters for style and content respectively. The following equations formulate the content and style losses:

\begin{equation}
\mathcal{L}_c(p)=\sum_{i\in{c}}\frac{1}{U_i}||\phi_i(p)-\phi_i(c)||_2^2
\end{equation}

\begin{equation}
\mathcal{L}_s(p)=\sum_{j\in{s}}\frac{1}{U_j}||G(\phi_j(p))-G(\phi_j(s))||_F^2
\end{equation}

where $\phi_l(p)$ represents the classifier activations at layer $l$, $U_l$ is the total number of units at layer $l$ and $G(\phi_l(p))$ is the Gram matrix associated with the layer $l$ activations.

\section{Experimental setup}
\label{sec: experiments}
Our evaluation methodology is based on testing the extrinsic object detection performance of oriented YOLO when trained with real+augmented data, with different mixing ratios of synthetic-to-real data, and tested on pure real data using mAP (mean average precision) as a performance metric. We test the following ratios synthetic:real; 1, 2, 4 and 8. Our first baseline is at 0 synthetic data augmentation, where oriented YOLO is trained only on real KITTI. The first set of experiments aims to show the advantage of synthetic data augmentation with different sensors models, and also with pure synthetic data augmentation without any sensors models mapping.


We also evaluate another extreme baseline where oriented YOLO is trained on pure synthetic CARLA data, with no real data at all. This is equivalent to the limiting case, when the simulated data dominates the real data. The test is performed as usual on real KITTI data. The aim of this experiment is two folds. First we want to see the discrepancy between the real and synthetic data distributions to prove the importance of using a realistic sensor model. Second, we want to test the extreme potential of synthetic data augmentation without any real data.

\subsection{Datasets}

All experiments are conducted on two datasets. For the real domain, KITTI dataset \cite{geiger2013vision} is used and for the simulated domain, data collected from CARLA simulator \cite{DBLP:journals/corr/abs-1711-03938} is used. Both datasets are in PCL 3D format. In order to adapt them to our task, 2D projection is performed on the data to produce Occupancy Grid Maps (OGMs). The projected data from both domains is of size $416 \times 416$. The same sensor setup as in KITTI dataset \cite{geiger2013vision} is used to configure the LiDAR sensor in CARLA simulator.

We have two kinds of data; one to train the sensors models, and the other to train oriented YOLO. From KITTI, 2000 and 12919 frames are used for neural sensor models and data augmentation training, respectively. The frames used for neural sensor models are included in the data augmentation training. As for simulated CARLA, 2000 frames are used for neural sensor models, however for data augmentation training, 12k, 25k, 50k and 100k frames are used for augmentation ratios: 1, 2, 4 and 8. The model is then tested on real 1400 BEV frames from KITTI dataset for all settings, using mAP (mean average precision) as a performance metric. In all our experiments we limit the classes to cars only because other classes like trucks and vans are not frequently encountered in CARLA scenarios, in contrast to KITTI which is more diverse, which would have created an issue of unbalanced data, if we used other classes.

\subsection{CycleGAN}

The basic architecture of the original paper \cite{Zhu-ICCV-2017} is used, we only adapted it to our data as illustrated in figure \ref{fig:vanilla_gan}.
The number of training samples is 2000 frames in each domain. $ \lambda $ is set to 50.0 as we noticed that increasing this parameter helps in reconstructing images in both directions leading to generating better images, this is because of using OGMs. Adam optimizer is used to train the network with learning rate of 0.0002 and batch size of 1. The network is trained for 200 epochs.

\subsection{Neural style transfer}
The network architecture used in this work is based on \cite{dumoulin2016learned}, but with smaller encoder of 4 convolution layers, each followed by a batch normalization layer, followed by a max pool layer, as mentioned before. This small network is pre-trained on a binary classification problem to classify between KITTI and CARLA. Accordingly, this binary classifier can capture the basic features of both domains. The full model is trained on the weighted summation of both content and style losses. We choose the $2nd$ layer activations for content loss and the activations from all $4$ layers of the feature extractor for style loss. Adam optimizer is used to train the network with learning rate of $1e-3$ and batch size of 4. The NST network is trained for around 40 epochs to achieve the desired results. The number of content frames is 2000, and the number of styles is 20, chosen to visually represent KITTI BEV. 46 different drives are selected from KITTI and clustered into 20 groups, each visually represents different surroundings and one frame is taken from each group.

\section{Discussion and results}
\label{sec: results}

As shown in the first row of table \ref{training-setup}, the mAP of oriented YOLO trained only on synthetic data  from CARLA (12K frames), and tested on real KITTI data (1.4K frames), is highly degraded to only 12\%. This observation proves the high discrepancy between the simulated and real data distributions. The next rows, shows the mAP with oriented YOLO trained on augmented data of KITTI+CARLA, with 100K synthetic CARLA frames. Compared to the baseline of 0 augmentation (second row), all other augmentation setups provide improvement ranging from 2\% in case of no NSM used to 8\% mAP in case of CycleGAN NSM. The marginal improvement of synthetic data augmentation is expected, and in fact proven by the low mAP of in the first row. 
\begin{table}
  \caption{Oriented YOLO mAP with and without data augmentation for different NSMs}
  \label{training-setup}
  \centering
  \begin{tabular}{l l}
    \toprule
    Training data                     & mAP (\%)      \\
    \midrule
    CARLA                     & 12.1  \\
    KITTI                     & 63.1  \\
    KITTI+100K(simulated)     & 65.3   \\
    KITTI+100K(NST)           & 69.3   \\
    KITTI+100K(CycleGAN)          & 71.5   \\
    \bottomrule
  \end{tabular}
\end{table}
\begin{figure}
  \centering
  \fbox{\includegraphics[width=0.6\linewidth]{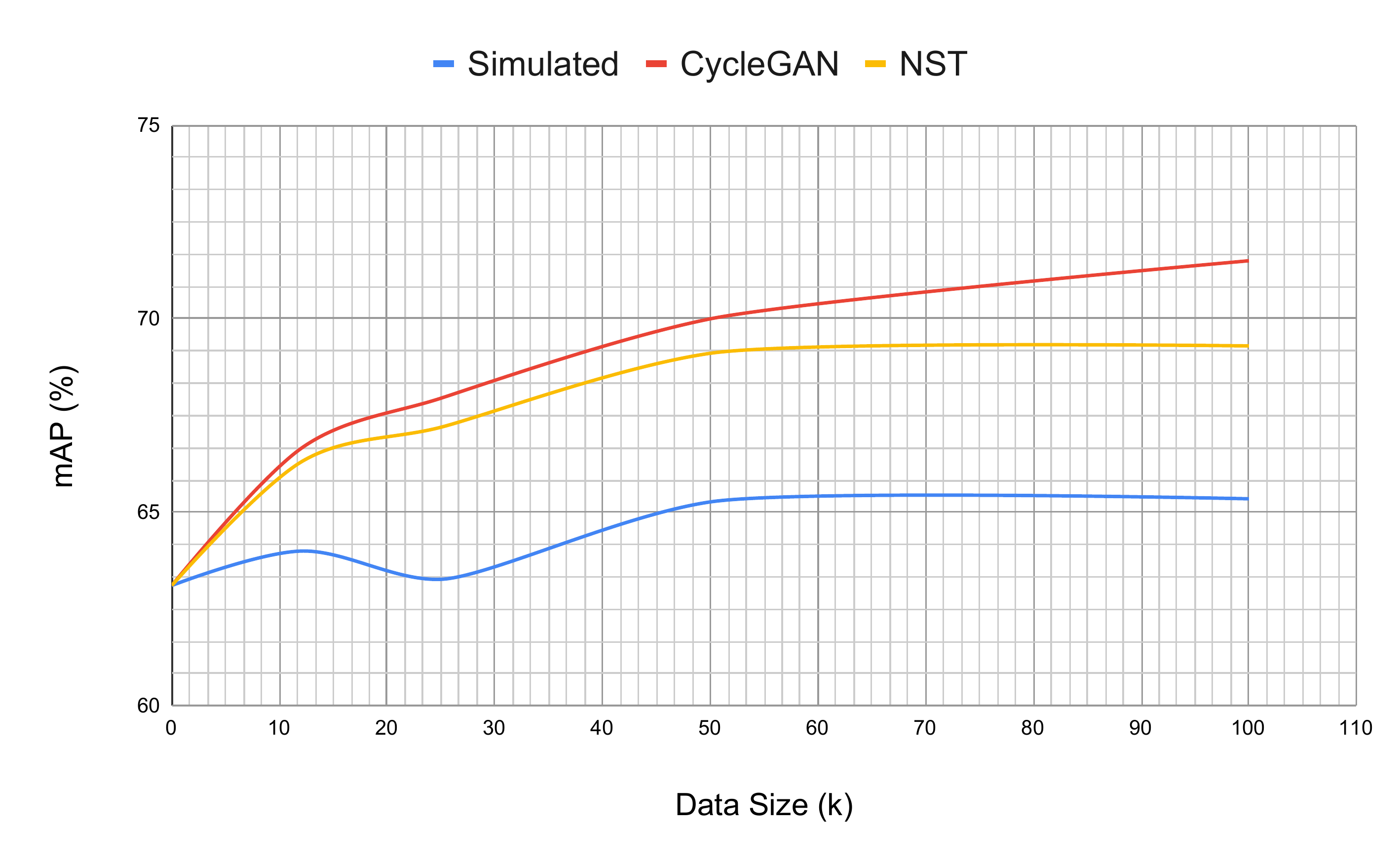}}  
  \caption{Results of increasing the amount of data augmentation and the corresponding mAP scores}
  \label{results_curve}
\end{figure}

This observation is more emphasised in figure \ref{results_curve}, where we plot the mAP at different amounts of data augmentation. The curve of simulated data fluctuates, and seems to saturate much faster than the other NSM curves. In fact, according to the result of pure synthetic data training, we can extrapolate that the synthetic data curve will decline to reach 12\% mAP in the limit when synthetic data dominates over real data (equivalent to the case in row 1 of table \ref{training-setup}). On the other hand, the other two curves of CycleGAN and NST show consistent increase in mAP with the increased amount of data augmentation, but are also projected to saturate. This saturation is also expected, due to the nature of the task, and also due to the network capacity, where we do not change the model with the increased amount of data. We expect that adapting the model size to increase with the increased amount of data will have the effect of delaying the saturation of the two NSM curves.

\begin{figure}
  \centering
  \begin{subfigure}[b]{.18\linewidth}
    \centering
    \includegraphics[width=\linewidth]{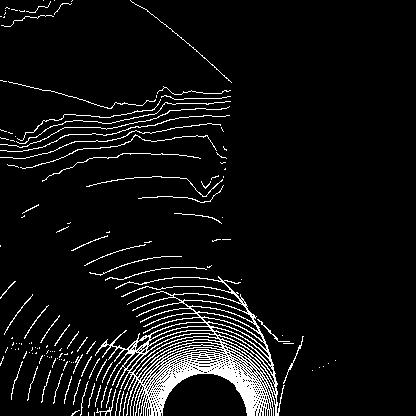}  
  \end{subfigure}
  \begin{subfigure}[b]{.18\linewidth}
    \centering
    \includegraphics[width=\linewidth]{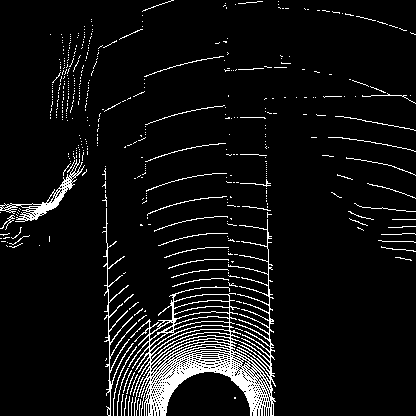}  
  \end{subfigure}
  \begin{subfigure}[b]{.18\linewidth}
    \centering
    \includegraphics[width=\linewidth]{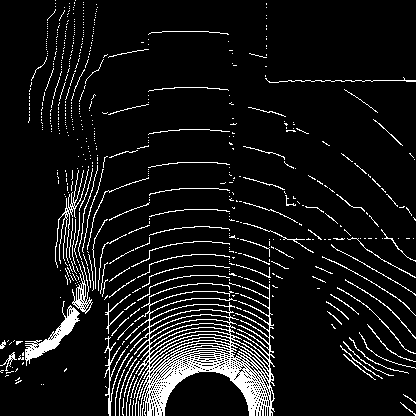} 
  \end{subfigure}
  \begin{subfigure}[b]{.18\linewidth}
    \centering
    \includegraphics[width=\linewidth]{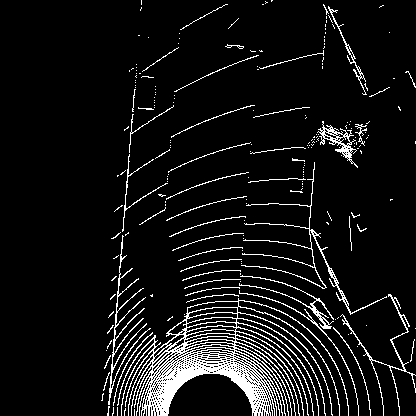} 
  \end{subfigure}\\
  \begin{subfigure}[b]{.18\linewidth}
    \centering
    \includegraphics[width=\linewidth]{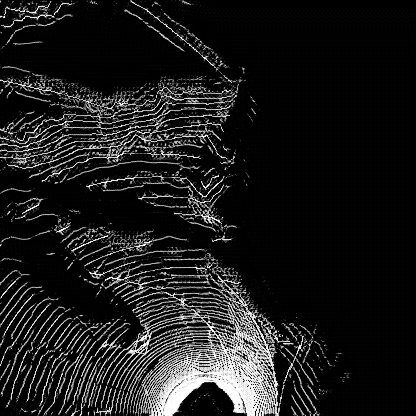}  
  \end{subfigure}
  \begin{subfigure}[b]{.18\linewidth}
    \centering
    \includegraphics[width=\linewidth]{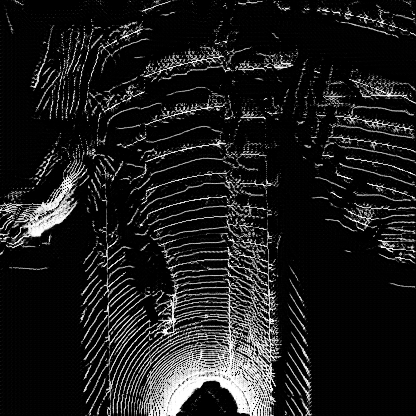}  
  \end{subfigure}
  \begin{subfigure}[b]{.18\linewidth}
    \centering
    \includegraphics[width=\linewidth]{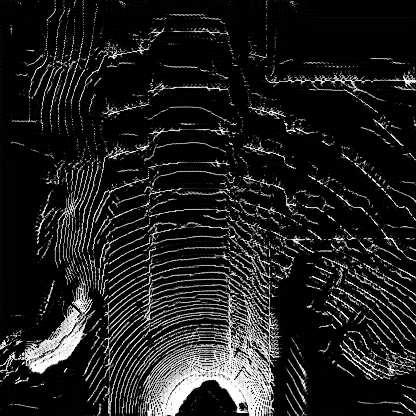} 
  \end{subfigure}
  \begin{subfigure}[b]{.18\linewidth}
    \centering
    \includegraphics[width=\linewidth]{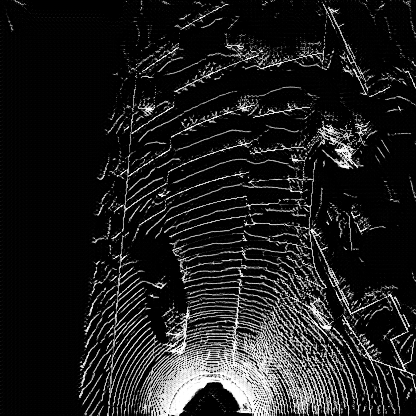} 
  \end{subfigure}\\
  \begin{subfigure}[b]{.18\linewidth}
    \centering
    \includegraphics[width=\linewidth]{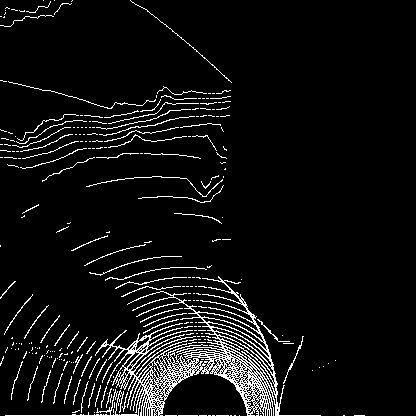}  
  \end{subfigure}
  \begin{subfigure}[b]{.18\linewidth}
    \centering
    \includegraphics[width=\linewidth]{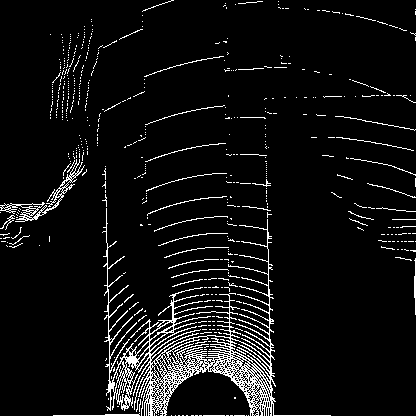}  
  \end{subfigure}
  \begin{subfigure}[b]{.18\linewidth}
    \centering
    \includegraphics[width=\linewidth]{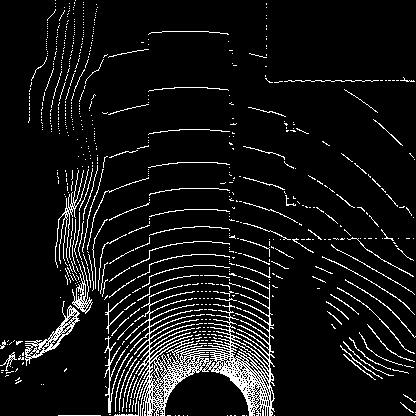} 
  \end{subfigure}
  \begin{subfigure}[b]{.18\linewidth}
    \centering
    \includegraphics[width=\linewidth]{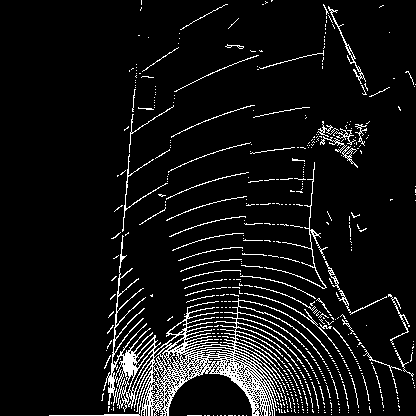} 
  \end{subfigure}
  \caption{Visual samples for neural sensors models. First row: CARLA simulated data. Second row: CycleGANs generated data. Third row: NST generated data}
  \label{visual_samples}
\end{figure}

The CycleGAN results are slightly better than the NST ones. Moreover, the CycleGAN provides a direct NSM, while the NST requires some workarounds to choose the styles of real data to use in generation. In figure \ref{visual_samples}, we show some samples from the 3 domains mappings. While the CycleGAN appears to model the noisy nature of real data, it also adds some artifacts. On the other hand, the NST seems more respecting the content as provided by CARLA, but not as good in modeling the noisy effects in the real KITTI data. The reason behind this could be the localized noise which does not affect the whole frame. This is also reflected in the curve in figure \ref{results_curve}, where we see that with small amounts of data augmentation, the CycleGAN and NST give almost the same mAP, because the ratio of synthetic data is still not dominant. As the ratio increases, CycleGAN begins to dominate, because it better models the real data, and hence the performance is improved on the real test data.

\section{Conclusion}
\label{sec: Conclusion}
In this work we proposed two NSMs for LiDAR data augmentation; CycleGAN and NST. Our evaluation included different ratios of data augmentation. The results show clear advantage of both sensor models over the baseline of augmenting raw simulated data of at least 6\% mAP improvement. Our experiments with increasing augmentation with NSM mapping show consistent increase in performance, although reaching saturation due to the nature of data and the capacity of the model, but at least no degradation is observed. CycleGAN provides slight improvement over NST in terms of mAP performance, and shows better visual results. Moreover, CycleGAN provides directly the NSM mapping from simulated to real domains, while NST requires workarounds via heuristics to feed the style with every frame generation.

\medskip

\bibliographystyle{unsrt}
\bibliography{ref}

\end{document}